\documentclass[prl, preprint, letter, showpacs, superscriptaddress]{revtex4}
\usepackage{amssymb}
\usepackage{amsmath}
\usepackage{bbm}
\usepackage{graphicx}
\usepackage{color}

\newcommand{\at}{\tilde\alpha}

\begin{document}

\author{Shenshen Wang}
\author{Peter G. Wolynes}
\affiliation{Department of Physics, Department of Chemistry and Biochemistry,  and Center for Theoretical Biological Physics, University of California, San Diego, La Jolla, CA 92093, USA}

\title{Effective temperature and glassy dynamics of active matter}
\date{\today}

\begin{abstract}

A systematic expansion of the many-body master equation for active matter, in which motors power configurational changes as in the cytoskeleton, is shown to yield a description of the steady state and responses in terms of an effective temperature. The effective temperature depends on the susceptibility of the motors and a Peclet number which measures their strength relative to thermal Brownian diffusion. The analytic prediction is shown to agree with previous numerical simulations and experiments. The mapping also establishes a description of aging in active matter that is also kinetically jammed.

\end{abstract}

\hyphenation{}

\maketitle



At the nanoscale and above we encounter active forms of matter in which internal or external energy sources supplement passive Brownian motion to allow large-scale structural rearrangements.
Constant agitation by biomolecular motors and force-generating polymerizations allows the cytoskeleton of eukaryotes to undergo adaptive dynamical and structural changes in response to environmental disturbances.
Other forms of active matter range from assemblies of entire microorganisms \cite{bacterial colonies} to collections of artificial microscopic swimmers which are self-propelled \cite{vibrated polar disks, sedimentation} and turbulently agitated active suspensions on the colloidal scale \cite{turbulent agitation}.


In this paper we show that, the steady-state statistical mechanics of these diverse out-of-equilibrium forms of active matter as well as their fluctuations and responses can be described using the concept of an effective temperature. The notion of effective temperature has been useful in describing passive glassy systems \cite{Cugliandolo et al}, weakly driven systems such as gently sheared supercooled liquids and glasses \cite{Barrat and Berthier,A J Liu} and vibrated granular matter \cite{vibrated granular}, as well as in approximate theories and simulations of active biological matter \cite{active processes1, active processes2, active processes3, active processes4}.

Modification of the fluctuation-dissipation theorem, signalling the nonequilibrium nature of the active processes like those in cells, has been directly observed in the mechanical properties of reconstituted cytoskeletal networks \cite{Mizuno_science, cytoplasmic diffusion}. These measurements were consistent with an effective temperature nearly $100$-fold higher than the ambient thermal temperature.


In this Letter we will show that the effective temperature for systems with motor-driven active processes can be directly related to the properties of the motors. We describe the motors as generating a time series of isotropic kicks leading to a master equation description for the dynamics. By investigating this master equation in the small kick limit we obtain an expression for $T_\textrm{eff}$ explicitly depending on the total motor activity and the susceptibility of the motor dynamics to imposed mechanical forces.
Motors lead to an enhanced diffusion for the active system regardless of susceptibility. Our theory predicts also a linear relation between the effective diffusion coefficient and the rescaled effective temperature, which depends on motor susceptibility. Our prediction is confirmed by recent numerical studies of systems with adamant motors \cite{Teff in active polymers}.

An important consequence of our analysis is that the effective temperature allows a simple rescaling of the equilibrium phase diagram to be used for the nonequilibrium motorized system. We show that the effective temperature coincides with the results based on previous self-consistent-field calculations \cite{JCP}. Active matter often is jammed and in a glassy state. The present analysis implies that Lubchenko-Wolynes aging theory \cite{aging} can be taken over to active matter simply by assigning the ambient temperature in that theory to be the motorized effective temperature and treating the fictive temperature as a history-dependent dynamically controlled variable as in structural glasses below their glass transition.



We start by following Shen and Wolynes \cite{TY} who modeled the stochastic nature of motor kicking via a master equation for the many-body probability distribution function $\Psi(\{\vec{r}\},t)$
\begin{equation}
\frac{\partial}{\partial t}\Psi(\{\vec{r}\},t)=(\hat{L}_\textrm{FP}+\hat{L}_\textrm{NE})\Psi(\{\vec r\},t).
\end{equation}
Here $\hat{L}_\textrm{FP}=D_0\sum_i\nabla_i\cdot\nabla_i-D_0\beta\sum_i\nabla_i\cdot(-\nabla_i U)$
is the usual many-body Fokker-Planck operator describing passive Brownian motion with $D_0$ denoting the thermal diffusion constant at bath temperature $T$ and $\beta=1/k_\textrm B T$. The gradient of the many-body interaction free energy function $U(\{\vec r\})=U(\vec{r}_1,\vec{r}_2,\cdots,\vec{r}_n)=\sum_{<ij>}u(\vec{r}_{i}-\vec{r}_j)$ gives the local force thus the thermal drift motion of individual particles and $\vec{r}_i$ is the position of the $i$th particle.
The effect of the nonequilibrium motor processes is summarized by an integral kernel
$\hat{L}_\textrm{NE}\Psi(\{\vec r\},t)=\int\Pi_i d\vec{r'_i}[K(\{\vec{r'}\}\rightarrow\{\vec{r}\})\Psi(\{\vec{r'}\},t)-K(\{\vec{r}\}\rightarrow\{\vec{r'}\})\Psi(\{\vec{r}\},t)]
$, where $K(\{\vec{r'}\}\rightarrow\{\vec{r}\})$ encodes the probability of transitions between different particle configurations.
The (motor) kicking noise is thus a finite jump process with a rate that depends on whether the free energy is increased or decreased by a motor step
\begin{equation}\label{kinetic rate}
k=\kappa[\Theta(\Delta U)\exp(-s_u\beta\Delta U)+\Theta(-\Delta U)\exp(-s_d\beta\Delta U)].
\end{equation}
Here $\Theta$ is the Heaviside step function and $\Delta U=U\left(\vec{r}+\vec{l}\right)-U\left(\vec{r}\right)$ is the energy change due to the kick identified by a vector $\vec l$. The kick step size $l$ and the basal kicking rate $\kappa$ define the dimensionless motor activity $\Delta:=\kappa\l^2/D_0$ which is analogous to the Peclet number in turbulent diffusion.
This model rate couples the chemical reactions leading to the motor activity to the local mechanical forces acting on the motor; the assumed dependence upon \textit{instantaneous} particle configuration reflects an assumed Markovian character of the dynamics without significant time delays.
This is an idealization of the biochemical mechanism of real motors that doubtless possess intermediates in their function.
We parametrize the coupling of a motor to the external forces it must overcome by the motor susceptibility $s$ which may take different values for uphill ($s_u$) moves and for downhill ($s_d$) moves. $s_u$ and $s_d$ depend on the chemical mechanism of the motors. When $s\rightarrow1$ the motors are susceptible; they slow down when they go up against mechanical obstacles and accelerate when they take energetically downhill steps; in contrast $s\rightarrow0$ corresponds to completely adamant motors which kick at an unperturbed rate being unaware of the instantaneous free energy landscape. Adamant motors use and waste a lot of energy.



Explicitly we write
\begin{widetext}
\begin{eqnarray}\label{NE full}
\hat{L}_\textrm{NE}\Psi(\{\vec{r}\},t)=\kappa\sum_{i}\int d\hat{n}\int d\vec{r'_i}
&\bigg\{&\delta(\vec{r}_i-\vec{r'}_i-\vec{l}\ )w[U(\cdots,\vec{r'}_i,\cdots)
-U(\cdots,\vec{r}_i,\cdots)]\Psi(\{\vec{r'}\},t)\nonumber\\
&-&\delta(\vec{r}_i-\vec{r'}_i+\vec{l}\ )w[U(\cdots,\vec{r}_i,\cdots)
-U(\cdots,\vec{r'}_i,\cdots)]\Psi(\{\vec{r}\},t)\bigg\}.
\end{eqnarray}
\end{widetext}
The angular integration denoted by $\int d\hat{n}$ averages over possible directions of kicking.
For simplicity we provide results for the isotropic case.
Again for simplicity we assume a fixed step length in the pair of delta functions.
Our description of the rates gives
$w[U_i-U_f]=\Theta(U_f-U_i)\exp[-s_u\beta(U_f-U_i)]+\Theta(U_i-U_f)\exp[-s_d\beta(U_f-U_i)]$.


When $s_u=s_d=s$, one finds even more simply
\begin{eqnarray}
\hat{L}_\textrm{NE}\Psi(\{\vec{r}\},t)=\kappa\sum_{i}\int d\hat{n}
\Big\{e^{-s\beta[U(\vec{r}_i)-U(\vec{r}_i-\vec{l})]}\nonumber\\
\times\Psi(\{\cdots,\vec{r'}_i=\vec{r}_i-\vec{l},\cdots\},t)\nonumber\\
-e^{-s\beta[U(\vec{r}_i+\vec{l})-U(\vec{r}_i)]}\Psi(\{\cdots,\vec{r}_i,\cdots\},t)\Big\}.
\end{eqnarray}
We assume the kicking of different particles at any time is uncorrelated.

To obtain our promised results, we expand the distribution function and the kinetic rate in powers of $\vec l$ up to the quadratic order. This immediately leads to an effective Fokker-Planck equation
\begin{equation}\label{detailed balance}
\frac{\partial}{\partial t}\Psi(\{\vec r\},t)=D_\textrm{eff}\sum_i\Big\{\nabla_i^2\Psi-\nabla_i\cdot\left[(-\nabla_i \beta_\textrm{eff}U)\Psi\right]\Big\},
\end{equation}
where
\begin{equation}\label{Deff}
D_\textrm{eff}=D_0\left(1+\frac{1}{2d}\frac{\kappa l^2}{D_0}\right),
\end{equation}
\begin{equation}\label{Teff}
\left(\beta_\textrm{eff}/\beta\right)^{-1}=T_\textrm{eff}/T=\left(1+\frac{1}{2d}\frac{\kappa l^2}{D_0}\right)\Big/\left(1+\frac{s}{d}\frac{\kappa l^2}{D_0}\right).
\end{equation}
Here the identity $\langle\cos^2\theta\rangle_{\hat n}=1/d$ is used for general spatial dimension $d$.
Rotational symmetry in motor susceptibility eliminates odd powers in $\vec l$ upon the angular integration.

These simple expressions (\ref{detailed balance})--(\ref{Teff}) valid for general dimensions have nontrivial implications.
First, in the small kick limit (kick step size is small compared to average particle separation yet the kicking rate can be quite high), the active system, while far from equilibrium in a strict sense, behaves as if it is at an effective equilibrium characterized by an effective temperature $T_\textrm{eff}$ which will however be quite different from the thermal temperature $T$. It follows that the steady state is described by the effective Boltzmann distribution as $\Psi(\{\vec{r}\})\varpropto\exp[-U(\{\vec{r}\})/k_\textrm BT_\textrm{eff}]$.
Second, the approach to effective equilibrium is governed by an effective diffusion constant $D_\textrm{eff}$
(Eq.~\ref{Deff}) which is enhanced by the active processes regardless of the motor adamancy.
This is consistent with recent experimental observation on enhanced cytoplasmic diffusion
which plays such an important role in cytoskeletal assembly \cite{cytoplasmic diffusion}.
Finally, it is clear from Eq.~(\ref{Teff}) that $T_\textrm{eff}$ can be predicted knowing only the motor activity $\Delta=\kappa l^2/D_0$ and the motor susceptibility $s$; susceptible motors with $s>1/2$ yield $T_\textrm{eff}<T$; more interestingly, for very high motor activity, i.e., $\Delta\gg1$, the effective temperature diverges as $T_\textrm{eff}/T\sim 1/(2s)$ as $s\rightarrow0$, indicating that intense kicking by adamant motors leads to a very high effective temperature just as observed in experiments and simulation studies.
In principle motors could have slip bonds \cite{slip bonds_exp, slip bonds_theory} leading to negative $s$ and negative effective temperatures, a situation we will explore in another publication.

The results are easily generalized to the case of asymmetric susceptibility. It is not difficult to show that
the factor $(s_u-s_d)$ accompanies all the cubic-and-above odd powers in $\vec l$ and thus doesn't modify the effective equilibrium at quadratic order.
Eqs.~(\ref{detailed balance})-(\ref{Teff}) remain intact except for a direct substitution of $(s_u+s_d)/2$ for $s$.
The effective temperature depends only on the \emph{sum} of uphill and downhill susceptibility, and if $s_u+s_d=1$ (i.e. $s=1/2$) we have $T_\textrm{eff}=T$.

\begin{figure}[htb]
\centerline{\includegraphics[angle=-90, scale=0.3]{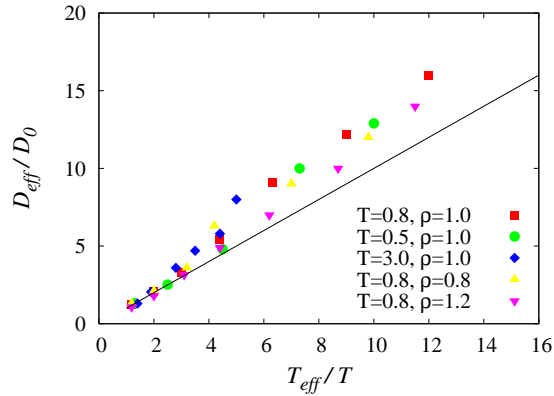}}
\caption{(color online). Rescaled diffusion constant versus rescaled effective temperature for a motorized semi-flexible polymer melt. The data points (colored symbols) were converted from Fig.~$18$ of Ref.~\cite{Teff in active polymers}. The solid line with slope $1$ gives the result predicted by the present theory.}
\label{Deff_vs_Teff}
\end{figure}

These predictions can be easily verified using results for various motile systems already studied numerically or experimentally.
A key prediction of our analysis is that $D_\textrm{eff}/D_0=T_\textrm{eff}/T$ for completely adamant motors with $s=0$. Numerical measurements of both the active diffusion constant and the effective temperature have been made by Loi \textit{et al.} \cite{Teff in active polymers} for an adamantly motorized semi-flexible polymer melt simulated using molecular dynamics techniques. Our theoretical prediction is plotted in Fig.~\ref{Deff_vs_Teff} along with their simulation results.
The data points nicely fall on the predicted straight line at low-$T_\textrm{eff}$ regime; even at relatively high $T_\textrm{eff}$ ($\geq4T$), only a modest deviation from the slope-$1$ line occurs which slowly increases with rising $T_\textrm{eff}$. The deviation is consistent with our scheme being perturbative in the kick size.
There are of course also larger error bars in numerical experiments for the larger $T_\textrm{eff}$ values.
It will be interesting to test our theory for partially susceptible motors experimentally.
Since $D_\textrm{eff}$ is independent of $s$, for a given motor activity $\Delta$, a higher susceptibility yields a lower $T_\textrm{eff}$ and thus a larger slope for the linear relation between $D_\textrm{eff}/D_0$ and $T_\textrm{eff}/T$.

Besides this relation our theory
suggests $T_\textrm{eff}$ should increase quadratically with the kick step size for adamant motors, just as is experimentally observed for the dependence on the Peclet number found both in sedimentation experiments \cite{sedimentation} for active colloidal suspensions under gravity, and in a numerical study of an active polymer melt \cite{Teff in active polymers}.

In an early study of the glassy dynamics of an assembly of motorized particles Shen and Wolynes \cite{TY} pictured the motors as introducing a modification to the Debye-Waller factors of the localized particles.
Their non-Hermitian variational approach to finding the steady state solution of the master equation turns out to be equivalent to closures for dynamic moments leading to an expression of the deviation of the total localization strength $\at$ from its mechanical value $\alpha$ in terms of the motor properties. On the other hand local mechanical feedback within the self-consistent phonon theory \cite{Fixman} gives back an $\alpha$ for the central particle that depends on the $\at$ of all its neighbors. Combining these two aspects allows a self-consistent determination of mean-field ($\alpha, \at$) solutions and identification of stability behavior accordingly.



Assuming $s_u=s_d=s$, the second moment closure given in \cite{TY} reduces to a simple expression
\begin{equation}\label{deviation}
\frac{\tilde\alpha-\alpha}{\tilde\alpha}=\left(s-\frac{1}{2}\right)\frac{\kappa l^2}{dD_0}\,e^{s(s-1)\alpha l^2}.
\end{equation}
Thus for $s=1/2$ ($T_\textrm{eff}=T$) chemical noise does not modify the mechanical stability ($\tilde\alpha=\alpha$); for $s<1/2$ ($T_\textrm{eff}>T$) stability is weakened ($\tilde\alpha<\alpha$) whereas for $s>1/2$ ($T_\textrm{eff}<T$) stability is enhanced ($\tilde\alpha>\alpha$).

\begin{figure}[htb]
\centerline{\includegraphics[angle=0, scale=0.4]{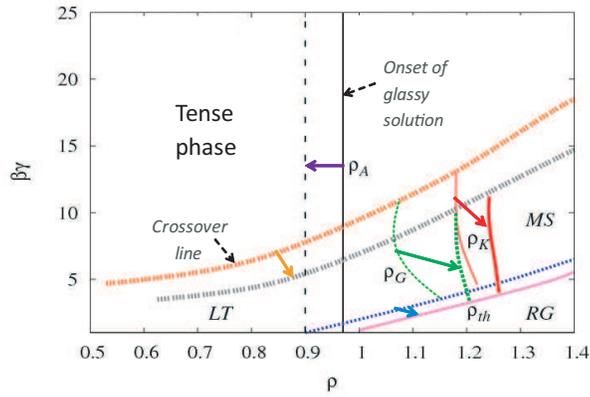}}
\caption{(color online). Superposed phase diagrams for the equilibrium and active cases for a cat's cradle with excluded volume having  $D_0=0.025\,\mu\textrm{m}^2\textrm{s}^{-1}$, $s=1$, $\kappa=10\,\textrm{s}^{-1}$, and $l=0.05\,\mu\textrm{m}$.
Parameters are comparable to those for a cytoskeleton. The motor kick size $l$ is much smaller than the typical crosslink separation which is on the order of $1$-$10\mu$m.
Solid arrows show the changes in characteristic behavior, pointing from the phase boundaries and glass transition densities for the equilibrium case to those for the active case. Above the upper phase boundary (the crossover line) the system is in the tense phase characterized by a smooth crossover from the loosely-tethered (LT) phase to the repulsive-glass (RG) phase as the crosslink density $\rho$ increases. Below the lower phase boundary (marked by $\rho_\textrm{th}$) the mobile phase becomes unstable and only the glassy state can be stabilized. Ref.~\cite{JCP} gives details of the analysis for the equilibrium case.}
\label{effective_equilibrium}
\end{figure}

Here we illustrate the connection to glassy dynamics by carrying out a self-consistent calculation (see Ref.~\cite{TY} for the detailed procedure) for a motorized version of the cytoskeletal network modeled as a ``cat's cradle" with excluded volume \cite{JCP}. Superimposing the phase diagrams for active and equilibrium cases over the same parameter ranges as shown in Fig.~\ref{effective_equilibrium} demonstrates that phase boundaries and characteristic densities scale just as we would predict from the kick length expansion.


The phase diagram shows diverse mechanical phases and possible transitions in between depending on the density ($\rho$) of the constituents and the effective stiffness ($\beta\gamma$) of the filamentous elements. In particular the coexistence of a loosely-tethered (LT) mobile phase and a repulsive-glass (RG) phase in the multiple-solution (MS) regime allows one to estimate the kinetic ($\rho_{_G}$) and thermodynamic ($\rho_{_K}$) glass transition densities by matching the free energies of the mobile and glassy states.


With both susceptible ($s=1$) and small-size ($l=0.05\mu\textrm{m}$) motor kicks, the variety of mechanical phases remains intact; but the phase boundaries shift due to the nonequilibrium effects.
The decrease in $\rho_{_A}$ leads to an enlarged stability region for the glassy state, while increases in glass transition densities $\rho_{_{G}}$ and $\rho_{_{K}}$ suggest that susceptible motor kicking can help resolve local constraints and thus allows a deeper descent into the energy landscape, leading to a denser packing where eventually structural rearrangements become too slow to be observed. The downward shift of both phase boundaries is simply captured by rescaling the equilibrium phase diagram with the inverse temperature ratio $\beta_\textrm{eff}/\beta$ given by Eq.~(\ref{Teff}). Adamant motor kicking gives rise to the opposite effect heating the system above its thermal temperature and reducing the glass transition density. Therefore equilibrium phase diagrams are still valid in the small kick limit yet with $T$ being replaced by $T_\textrm{eff}$.

The fact that the master equation precisely reduces to a Fokker-Planck equation allows us to apply the effective temperature to analyze the dynamics/kinetics of systems that are structurally out-of-equilibrium, i.e., jammed.
The slow relaxation of the cytoskeletal mechanics \cite{cytoskeletal slow dynamics} thus is related to glassy behavior in the aging regime.
Quantitatively, jammed cytoskeletal systems should be described by the aging theory for structural glasses developed by Lubchenko and Wolynes \cite{aging}.
It is only necessary to replace in their theory the absolute ambient temperature which describes vibrations in glasses by the effective temperature of the motors.
The fictive temperature depends then on the history of the cytoskeleton motor activity and preparation.
This mapping should make it possible to predict the dependence on motor properties of the rheology of active matter.



Support from the Center for Theoretical Biological Physics sponsored by the NSF (Grant PHY-0822283) is gratefully acknowledged.

\end{document}